\documentclass[twocolumn,floatfix,showpacs,amsmath,amssymb,prl]{revtex4}
\usepackage{times}
\usepackage{graphicx}
\usepackage{dcolumn}
\usepackage{bm}

\begin{document}

\title{High light intensity photoassociation in a Bose-Einstein condensate}
\author{Thomas Gasenzer}
\affiliation{Institut f\"ur Theoretische Physik, Philosophenweg 16, 
69120 Heidelberg, Germany}
\date{\today}

\begin{abstract} 
We investigate theoretically the molecular yield in photoassociation of Bose-Einstein condensed sodium atoms for light intensities of the order of and above those applied in a recent experiment. 
Our results show that the rate at which ground state molecules may be formed saturates at high light intensities whereas the loss rate of condensate atoms does not.
This is caused by the opposing roles of the short and long range pair correlations present near resonance under the influence of the laser and is crucial for the development of efficient photoassociation procedures in a condensate.
\end{abstract}
\pacs{03.75.Nt, 03.75.Kk, 33.70.-w, 05.30.-d\hfill HD--THEP--04--02}
\maketitle
The search for efficient ways for photoassociation (PA) \cite{Weiner99} of Bose-Einstein condensed atoms to condensed diatomic molecules has gained much inertia in the recent past.
As with PA in non-degenerate gases (cf.~\cite{Prodan03} and refs.~therein) growing theoretical \cite{Bohn96,Kostrun00} and experimental \cite{McKenzie02,Prodan03} attention has been paid to the investigation of possible limits to the achievable PA rate in condensates.
Of particular interest is the question at which laser intensities $I$ the rate, which is predicted to increase linearly with $I$ in the low intensity limit, saturates.

In the experiment reported in Ref.~\cite{McKenzie02} a single color laser light pulse has been applied such that the atomic collisions occur close to a photoassociative scattering resonance.
From the measured fraction of atoms remaining after the pulse on-resonance loss rates have been deduced for various peak intensities of the light.  
These loss rates were found to agree well with the two body theory of atoms scattering under the influence of a near resonant photoassociative light field \cite{Bohn96}.
In particular, the experimental results for the time evolution of the atom density follow, to a good accuracy, the rate equation $\dot{n}(t)=-K\,n^2(t)$ if, on resonance, the rate constant $K$ is taken to be $K_0=(8\pi\hbar/m)|\Im a_0|$ where $a_0$ is the on-resonance complex scattering length derived from the scattering matrix \cite{Bohn96,McKenzie02}.

Recent many body calculations \cite{Kostrun00} have suggested an upper limit to the rate of condensate loss due to photoassociation of $K_{\rm PA}\simeq (2\pi\hbar/m)\ell$, where $\ell=n^{-1/3}/2\pi$ is set by the mean atomic distance $n^{-1/3}$.
However, it has been pointed out in Ref.~\cite{McKenzie02} that for the maximum atomic density used in the experiment one has $(h/m)\ell=3.8\cdot10^{-10}\,$cm$^3$s$^{-1}$ which is of the order of the PA rate constant $K_0$ reached at the highest applied light intensities.
The dependence of the measured $K_0(I)$ so far gives no evidence for a saturation at higher intensities $I$ \cite{McKenzie02,LettPrivComm03} and below the unitarity limit \cite{Prodan03}.

In this article we show that while indeed the formation rate for {\em ground state} molecules saturates for large $I$ the rate of loss from the condensate does not, due to the formation of non-condensed atom pairs.
The theory applied in the present work is described in detail in Refs.~\cite{KB02,KGB03,KGJB03} and fully takes into account the direct loss of condensate atoms into non-condensed, motionally excited states, the production and dissociation of diatomic molecules during the laser assisted collisions, as well as the spontaneous decay of electronically excited atoms.

In the experiment \cite{McKenzie02}, a $\lambda=591\,$nm laser couples the open channel of electronic $|F=1,m_F=-1\rangle$ ground state $^{23}$Na atoms, colliding at near threshold energies, to the ($J=1,\nu=135$) rovibrational level of the closed channel excimer potential of atom pairs which asymptotically are in the electronic states $3^2$S$_{1/2}$ + $3^2$P$_{1/2}$ (cf.~Fig.~1 in Ref.~\cite{McKenzie02}).
A rectangular intensity profile was chosen, with ramp and fall times shorter than $0.5\,\mu$s and a length of up to $400\,\mu$s.
In the absence of the laser the collisions of the ground state atoms in the $\{-1,-1\}$ open channel are characterized by the background potential $V_{\rm bg}(r)$.
The laser couples the atoms in this channel to a particular rovibrational state $|\phi_\nu\rangle$ of the closed channel excimer potential $V_{\rm cl}(r)$ (cf.~Fig.~1 in Ref.~\cite{McKenzie02}).
In the absence of Doppler broadening the coupling of the open channel to other closed channel states may be neglected.
Also coupling to other open channels leading, e.g., to inelastic two body loss, is sufficiently weak.
In the interaction picture and the rotating wave approximation (rwa) we may therefore reduce the Hamiltonian of untrapped interacting atom pairs to
\begin{eqnarray}
 \label{H2B}
  H_\mathrm{2B}
  &=&-\frac{\hbar^2\nabla^2}{m}
  |\mathrm{bg}\rangle\langle\mathrm{bg}|
  +|\phi_\nu,{\rm cl}\rangle (E_\nu-\hbar\omega) \langle\tilde\phi_\nu,{\rm cl}|
  \nonumber\\
 \label{V2B}
  &&+\ (|\mathrm{bg}\rangle,|\mathrm{cl}\rangle)
      \left(\begin{array}{cc}
      V_{\rm bg} & 
      WP_\nu \\
      P_\nu W  & 
      0
      \end{array}\right)
      \left(\!\!\begin{array}{c}
      \langle\mathrm{bg}|\\ 
      \langle\mathrm{cl}|
      \end{array}\!\!\right).
\end{eqnarray}
Here $|\mathrm{bg}\rangle=|1,-1\rangle|1,-1\rangle$ is the product of hyperfine states associated with the $\{-1,-1\}$ open channel and $|\mathrm{cl}\rangle$ is the superposition of products of atomic states associated with the closed channel.
The closed channel Hamiltonian $H_{\rm cl}(r)=-\hbar^2\nabla^2/m+V_{\rm cl}(r)$ has been replaced by the operator $E_\nu P_\nu$, with the (quasi \cite{fn:lefteigenstates}) projector $P_\nu=|\phi_\nu\rangle\langle\tilde\phi_\nu|$ onto the resonantly coupled excimer state, $H_{\rm cl}(r)\phi_\nu(r)=E_\nu\phi_\nu(r)$.
The spontaneous decay of $|\phi_\nu\rangle$ to hot atom pairs or ground state molecules is taken into account in the imaginary part of $E_\nu=\hbar(\omega_\nu-i\gamma/2)$ given by the measured lifetime $\tau=1/\gamma=8.6\,$ns \cite{McKenzie02}.
$\omega=2\pi c/\lambda$ is the laser frequency.
The laser induced coupling is described by $W(\mathbf{x},t)=(\hbar/2)\Omega_0(t)\exp[i(\mathbf{k}\mathbf{x}-\omega t)]$, with Rabi frequency $\Omega_0(t)$ whose time dependence takes into account the envelope of the light field.
In Eq.~(\ref{H2B}) the trapping potential may be neglected since it does not vary on the scale of the binary interactions.

The microscopic many body approach \cite{KB02,KGB03} includes the dynamics of the mean field $\Psi(\mathbf{x},t)=\langle \psi_{\rm g}(\mathbf{x}) \rangle_t$ as well as of the density matrix of the non-condensed fraction $\Gamma_{ij}(\mathbf{x},\mathbf{y},t)=\langle
\psi_j^{\dagger}(\mathbf{y}) \psi_i(\mathbf{x}) \rangle_t - 
\Psi^*(\mathbf{y},t)\Psi(\mathbf{x},t)$ $\delta_{i\rm g}\delta_{j\rm g}$ and the pair correlation function $\Phi_{ij}(\mathbf{x},\mathbf{y},t)$ $=\langle
\psi_j(\mathbf{y}) \psi_i(\mathbf{x}) \rangle_t - 
\Psi(\mathbf{y},t) \Psi(\mathbf{x},t)\delta_{i\rm g}\delta_{j\rm g}$ of the bosonic field operators for both, atoms asymptotically in the ground and excited electronic states ($i,j\in\{$g,e$\}$). 
All physical quantities of the atom gas may be described in terms of correlation functions the dynamics of which is determined by an infinite hierarchy of coupled differential equations.
As shown in Refs.~\cite{KB02,KGB03} this exact hierarchy can be truncated in a consistent manner. 
Within the two-channel description introduced above the mean field of the ground state condensate atoms and and the pair functions are, to leading order, determined by the coupled equations:
\begin{widetext}
\begin{align}
 \label{meanfield}
  \nonumber\\[-1cm]
  i\hbar\frac{\partial}{\partial t}\Psi(\mathbf{x},t)
  = H_{\rm g}^\mathrm{1B}(\mathbf{x})
  \Psi(\mathbf{x},t)
  +\!\!\int\! d\mathbf{y}\Psi^*\!(\mathbf{y},t)\Big(
  V_{\rm bg}(|\mathbf{x}\!-\!\mathbf{y}|)
  \Big[
    \Phi_{\rm bg}(\mathbf{x},&\mathbf{y},t)+
    \Psi(\mathbf{x},t)\Psi(\mathbf{y},t)
    \Big]
    \!+\!\hbar\Omega(\mathbf{x}\!-\!\mathbf{y},t)\Phi_{\rm cl}(\mathbf{x},\mathbf{y},t)\Big),\\
 \label{pairfunction}
  i\hbar\frac{\partial}{\partial t}
  \left(\begin{array}{c}
        \Phi_{\rm bg}(\mathbf{x},\mathbf{y},t)\\
	\Phi_{\rm cl}(\mathbf{x},\mathbf{y},t)
  \end{array}\right)
  = H^\mathrm{2B}(\mathbf{x},\mathbf{y},t)
  \left(\begin{array}{c}
        \Phi_{\rm bg}(\mathbf{x},\mathbf{y},t)\\
	\Phi_{\rm cl}(\mathbf{x},\mathbf{y},t)
  \end{array}\right)+
  &\left(\begin{array}{c}
        V_{\rm bg}(|\mathbf{x}-\mathbf{y}|)\\
	\hbar\Omega(\mathbf{x}-\mathbf{y},t)
  \end{array}\right)
  \Psi(\mathbf{x},t)\Psi(\mathbf{y},t).
\end{align}
\end{widetext}
Here $H^\mathrm{1B}_{\rm g}(\mathbf{x})=-\hbar^2\nabla^2/2m+V_\mathrm{trap}(\mathbf{x})$ is the Hamiltonian of a single atom in the electronic ground state.
$\Omega(\mathbf{x},t)=\Omega_0(t)\cos(\omega t-\mathbf{k}\mathbf{x})$ describes the laser field.
Furthermore $H^\mathrm{2B}(\mathbf{x},\mathbf{y},t)$ is the general two channel two-body Hamiltonian of interacting atom pairs which in the interaction picture and rwa has the form given in Eq.~(\ref{H2B}). 
Inelastic loss phenomena due to e.g.~three body recombination \cite{KB02,Koehler02} will be neglected here.

Using Eq.~(\ref{H2B}) the closed system of Eqs.~(\ref{meanfield}), (\ref{pairfunction}) may be solved to obtain a single non-Markovian non-linear Schr\"o\-din\-ger equation for the mean field of the condensed electronic ground state atoms \cite{KB02,KGB03,GKGJT04}:
\begin{align}
  \label{NMNLSE}
  \nonumber
  i\hbar\frac{\partial}{\partial t}\Psi(\mathbf{x},t)
  \ =\ & 
  H^\mathrm{1B}_{\rm g}(\mathbf{x})
  \Psi(\mathbf{x},t)\\ 
  -\ &\Psi^*(\mathbf{x},t)\int_{t_0}^\infty d\tau
  \Psi^2(\mathbf{x},\tau)\frac{\partial}{\partial \tau}h(t,\tau),
\end{align}
where  
$h(t,\tau)=(2\pi\hbar)^3\langle0,\mathrm{bg}|V_\mathrm{2B}(t)
U_\mathrm{2B}(t,\tau)|0,\mathrm{bg}\rangle\,\theta(t-\tau)$. 
$V_\mathrm{2B}$ is given by the third term in Eq.~(\ref{H2B}), $U_\mathrm{2B}(t,\tau)$ is the time evolution operator corresponding to the Hamiltonian (\ref{H2B}), $\theta(t-\tau)$ is the step function that evaluates to zero except for $t>\tau$, and $|0,\mathrm{bg}\rangle$ is the zero momentum plane wave state of the relative motion of two atoms in the open channel.
In deriving Eq.~(\ref{NMNLSE}) we have assumed that at the initial time 
$t_0$ the gas is effectively uncorrelated, e.g., a weakly interacting ground state Bose-Einstein condensate, fully described by $\Psi(\mathbf{x},t_0)$. 

At the low relative kinetic energies in a condensate the binary scattering properties in the absence of laser light are accurately determined by the scattering length $a_{\rm bg}$ as well as by the energy of the last bound state of $V_{\rm bg}$ \cite{Gao98}. 
For the $^{23}$Na atoms in the $\{-1,-1\}$ channel we use the theoretically determined value of $a_{\rm bg}=54.9\,a_{\rm Bohr}$ ($a_{\rm Bohr}=0.052918\,$nm) \cite{Mies02}.
The energy of the last bound state (vibrational quantum number $\nu=65$) has been measured to be $E_{\rm bg}/h=317.78\,$MHz \cite{Samuelis00}.
Any model of $V_{\rm bg}$ which properly accounts for $a_{\rm bg}$ and $E_{\rm bg}$ will recover the correct low energy scattering properties.
To solve Eq.~(\ref{NMNLSE}) we apply the minimal model introduced in Ref.~\cite{KGB03}: $V_{\rm bg}=|\chi_{\rm bg}\rangle\xi_{\rm bg}\langle\chi_{\rm bg}|$, with $m\xi_{\rm bg}/(4\pi\hbar^2)=-191.4\,$a$_{\rm Bohr}$ and the convenient Gaussian function $\chi_{\rm bg}(r)=\exp(-r^2/2\sigma_{\rm bg}^2)/(\sqrt{2\pi}\sigma_{\rm bg})^3$, $\sigma_{\rm bg}=23.97\,$a$_{\rm Bohr}$.

A similar ansatz is chosen for the coupling: $\frac{1}{2}\Omega_0|\phi_\nu\rangle=|\chi\rangle\xi$, with $\xi=-(2\pi)57\,$GHz $(a_{\rm Bohr}^3/($kW$/$cm$^2))^{1/2}$ and $\chi(r)=\exp(-r^2/2\sigma^2)/(\sqrt{2\pi}\sigma)^3$, $\sigma=3.55\,a_{\rm Bohr}$.
Using this parametrization the zero momentum limit of the scattering amplitude derived from Eq.~(\ref{H2B}) is equal to $4\pi\hbar^2a(\Delta,I)/m$, with $a$ given by (cf.~\cite{Bohn96})
\begin{equation}
  \label{aDelta}
  a(\Delta,I)
  = a_{\rm bg} -\lim_{k\to0}\frac{\Gamma(I)/2k}
                                 {\Delta+\delta_\nu(I)-i\gamma/2},
\end{equation}
where $\Delta={\rm Re}(E_\nu)/\hbar-\omega$ is the detuning of the laser frequency $\omega$ from the transition energy.
$\Gamma(I)$ is the width of the laser induced transition, depending on the light intensity $I$ and, for small $k$, linearly on the relative momentum $\hbar k$ of the colliding atoms \cite{Bohn96}.
We use the theoretical value of $\partial(\Gamma(I)/k)/\partial I=0.69743\,$ms$^{-1}($kW/cm$)^{-1}$ \cite{McKenzie02}.
$\delta_\nu(I)$ is the intensity dependent shift of the resonance which experimentally has been found to depend linearly on $I$, $\partial\delta_\nu(I)/\partial I=-164\,$MHz$($kW/cm$^2)^{-1}$ \cite{McKenzie02}.

\begin{figure}[tb]
\begin{center}
\includegraphics[width=0.45\textwidth]{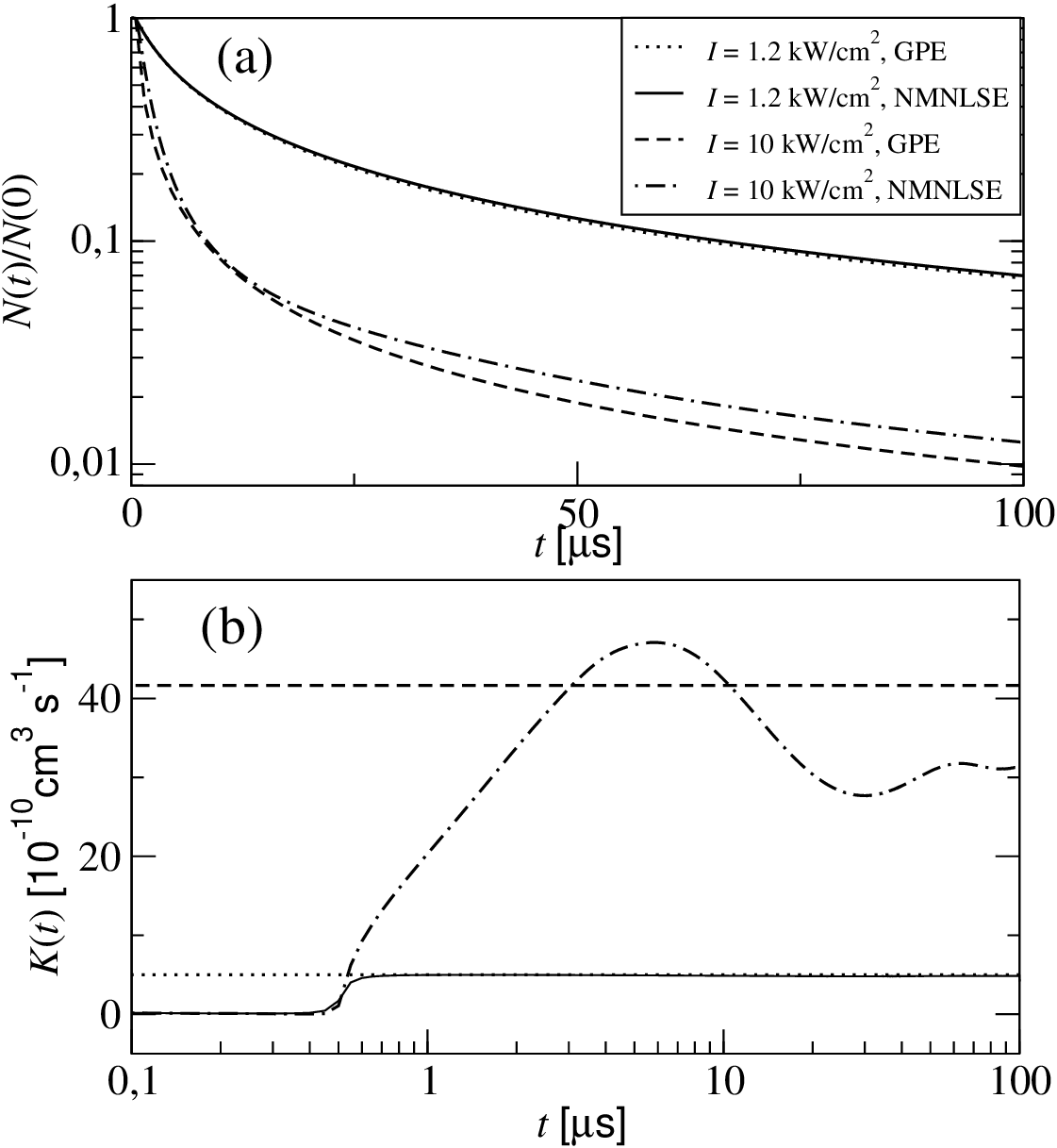}
\end{center}
\vspace*{-3ex}
\caption{
(a) Time evolution of the remaining fraction of the initial atom number $N(0)\!\!=\,$4.0$\,\cdot\,$10$^{6}$ during a $100\,\mu$s pulse, with peak intensities of $1.2\,$kW/cm$^2$ (solid) and $10\,$kW/cm$^2$ (dash-dotted line). 
The fraction $N(t)/N(0)$ is given on a logarithmic scale.
Shown are the solutions of the non-Markovian non-linear Schr\"odinger equation (NMNLSE) (\ref{NMNLSE}) and, for comparison, the evolution according to the Gross-Pitaevskii equation (GPE),  $n(\mathbf{x},t)=n(\mathbf{x},0)[1+K_0n(\mathbf{x},0)(t-t_{\rm rise})]^{-1}$, $t_{\rm rise}=0.5\,\mu$s, with $K_0=(8\pi\hbar/m)|\Im a(\Delta,I)|$.
(b) Time evolution of the temporally local decay `rate' $K(t)=-\dot{N}(t)/\int d\mathbf{x}\,n^2(\mathbf{x},t)$ correspoding to the cases considered in (a).
Time is shown on a logarithmic scale.\\[-0.8cm]
}
\label{fig:condtimedep}
\label{fig:locdecayrate}
\end{figure}

In Fig.~\ref{fig:condtimedep}a the time evolution of the condensate fraction according to Eq.~(\ref{NMNLSE}) is shown for a $100\,\mu$s light pulse of intensity $I(t)$ which is ramped, between $t=0$ and $t=0.5\,\mu$s, linearly to its peak value $I_{\rm max}=1.2\,$kW cm$^{-2}$ (dash-dotted line), the maximum intensity reported in Ref.~\cite{McKenzie02}.
The detuning $\Delta=(2\pi)\,196.8\,$MHz is chosen such that it compensates for $\delta_\nu(I_{\rm max})$ in Eq.~(\ref{aDelta}). 
The initial number of atoms in the spherical harmonic trap with frequency $\nu_{\rm ho}=198\,$Hz is $N(0)=4.0\cdot10^{6}$.
This is compared to the time evolution with a constant scattering length and loss rate, shown as the dotted line in Fig.~\ref{fig:condtimedep}a, where the local density evolves according to the Gross-Pitaevskii equation (GPE), as $n(\mathbf{x},t)=n(\mathbf{x},0)[1+K(\Delta,I_{\rm max})n(\mathbf{x},0)t]^{-1}$, with $K(\Delta,I)=(4h/m)|\Im a(\Delta,I)|$.

In Fig.~\ref{fig:condtimedep}a we also compare the solutions of Eq.~(\ref{NMNLSE}) and of the GPE for $I_{\rm max}=\,$10$\,$kW/cm$^{2}$.
The essential difference between the dashed and dash-dotted loss curves becomes apparent when considering the temporally local decay `rate' $K(t)=-\dot{N}(t)/\int d\mathbf{x}\,n^2(\mathbf{x},t)$ shown in Fig.~\ref{fig:locdecayrate}b.
While, after the 100$\,\mu$s length of the pulse the final densities are comparable, there is a fundamental difference in the amount of condensate loss at a certain time.
This underlines the significance of the non-Markovian character of the dynamic equation (\ref{NMNLSE}) and shows that the condensate loss at high intensities 
may not be described by a simple rate equation.
\begin{figure}[tb]
\begin{center}
\includegraphics[width=0.45\textwidth]{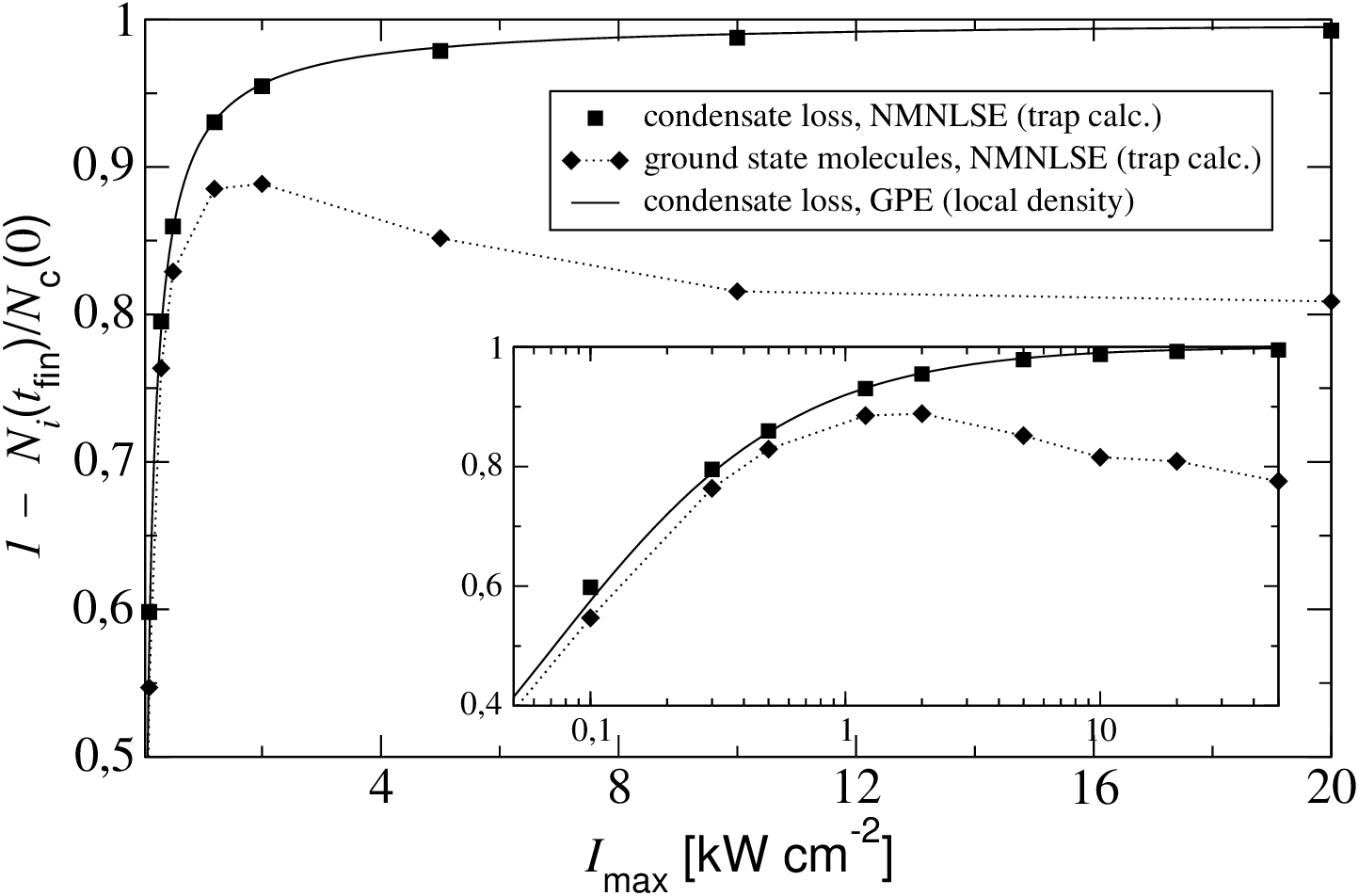}
\end{center}
\vspace*{-3ex}
\caption{
Loss fractions $1-N_{i}(t_{\rm fin})/N_{\rm c}(0)$, $i=$ `c', `tot', of the condensate (filled squares) and total atom (filled diamonds) loss at the end of the laser pulse, $t_{\rm fin}=\,$100$\,\mu$s, for different laser intensities $I_{\rm max}$.
The fraction indicated by the diamonds corresponds to the {\em number of ground state molecules} formed.
The solid line shows the result of a solution of the GPE in local density approximation.
The inset shows the same quantities on a logarithmic scale.
}
\label{fig:NlossofI}
\end{figure}

One sees that the many body theory which takes into account all pair correlations created by elastic two body collisions does not give a saturation of the condensate loss rate above ca.~$I=1\,$kW cm$^{-2}$.
Fig.~\ref{fig:NlossofI} shows, for different $I_{\rm max}$, the fraction of atoms lost from the condensate (filled squares) at the final time $t_{\rm fin}=\,$100$\,\mu$s of the light pulse.
For large $I_{\rm max}$ the loss approaches 100$\%$.
This loss is compared to the on resonance decrease of the total number of atoms $N_{\rm tot}=\int d\mathbf{x}[|\Psi(\mathbf{x})|^2+\Gamma(\mathbf{x},\mathbf{x})]$ due to the spontaneous decay of $|\phi_\nu\rangle$ (filled diamonds), calculated from $\hbar\dot{N}_{\rm tot}=-\gamma\int d\mathbf{R}\,|\int d\mathbf{r}\,\tilde\phi_\nu^*(\mathbf{r})\Phi_{\rm cl}(\mathbf{R},\mathbf{r},t)|^2$ \cite{G04}, $\mathbf{R}$ and $\mathbf{r}$ being the c.m.~and relative coordinates.
Since in the 2-channel picture considered here ground state molecules can only be formed through this decay the upper limit seen in Fig.~\ref{fig:NlossofI} implies a {\em limitation of the molecular yield} for large $I_{\rm max}$.

Which are the physical reasons behind these limits?
Depending on the magnitude of $\Delta$, during the ramps the laser causes the closed channel bound state energy to closely approach or cross the open channel threshold.
A rapid ramp of $a$ from $a(\Delta,0)=a_{\rm bg}$ to a particular value near resonance represents a strong sudden change in the binary interactions.
The colliding condensate atoms are converted both, into molecules and directly to excited correlated pairs which no longer belong to the condensate \cite{KGB03,KGJB03,GKGJT04}. 

\begin{figure}[tb]
\begin{center}
\includegraphics[width=0.45\textwidth]{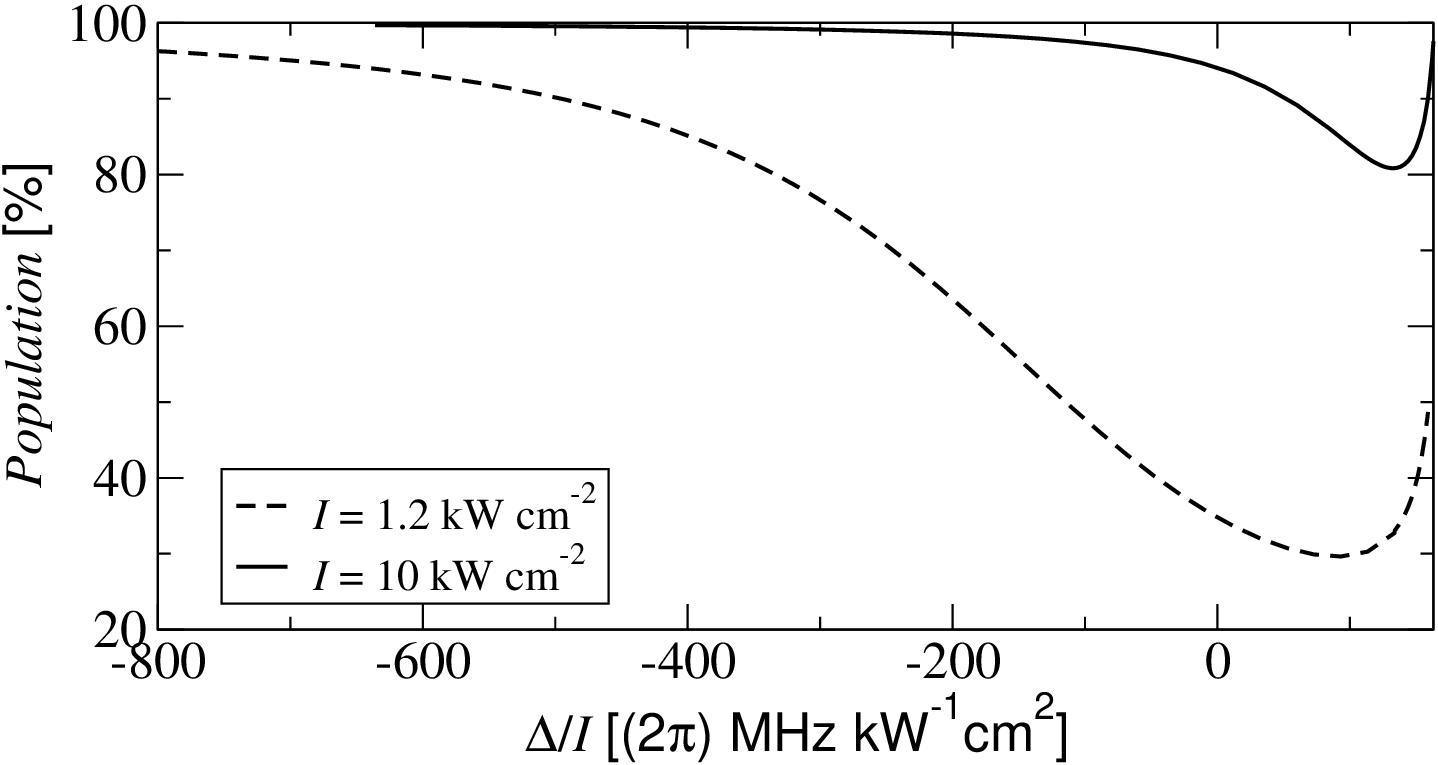}
\end{center}
\vspace*{-3ex}
\caption{
Population $({\cal N}_{\rm d}^2-1)/{\cal N}_{\rm d}^2$ of the open channel component of the dressed state wave function as a function of the renormalized detuning $\Delta/I$ from the resonance, for $I=\,$1.2$\,$kW/cm$^2$ (dashed line) and $I=\,$10.0$\,$kW/cm$^2$ (solid line).
The real part of the dressed state energy vanishes at $\Delta=($2$\pi)\,$160$\,$MHz and $\Delta=($2$\pi)\,$163$\,$MHz, respectively. Due to the finite lifetime of these states the populations are reduced to ca.~50$\%$ and 98$\%$, respectively, as compared to 100$\%$ in the case that the closed channel state is stable, $\gamma=0$.\\[-0.8cm]
}
\label{fig:bgfractionofphib}
\end{figure}
\begin{figure}[tb]
\begin{center}
\includegraphics[width=0.45\textwidth]{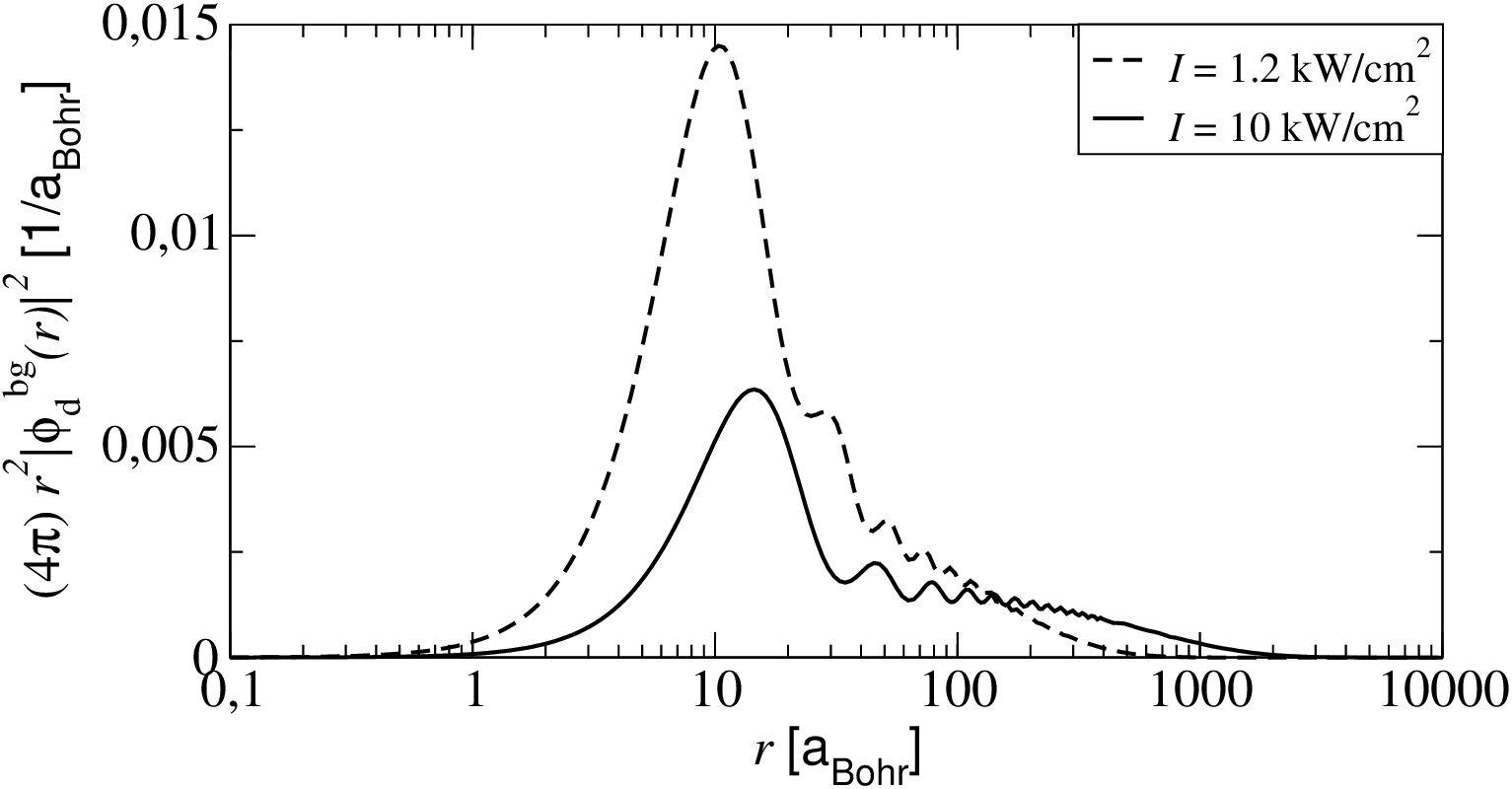}
\end{center}
\vspace*{-3ex}
\caption{
Open channel component of the radial dressed state density for $I=\,$1.2$\,$kW/cm$^2$ (dashed line) and $I=\,$10$\,$kW/cm$^2$ (solid line), and a detuning $\Delta=($2$\pi)\,$192.3$\,$MHz and $\Delta=($2$\pi)\,$1631.2$\,$MHz, respectively, which is chosen such that Re$E_{\rm d}$ is just below threshold.
$r$ is given on a logarithmic scale.
The wave functions have been computed using the minimal ansatz for $V_{\rm bg}$ and $\Omega_0|\phi_\nu\rangle$ which agrees with a complete coupled channels state at radii large compared to the van der Waals length $\frac{1}{2}(mC_6/\hbar^2)^{1/4}=\,$45$\,a_{\rm Bohr}$ \protect\cite{KGJB03}.
The extension in $r$ scales with the modulus of the scattering length: $|a($192.3$\,$MHz$,$1.2$\,$kW/cm$^2)|=\,$161$\,a_{\rm Bohr}$, $|a($1631$\,$MHz$,$ 10$\,$kW/cm$^2)|=\,$863$\,a_{\rm Bohr}$.
(Cf.~also Ref.~\protect\cite{Deb03}.)\\[-0.7cm]
}
\label{fig:phibbgofr}
\end{figure}

Let us consider the negative energy dressed state wave function of the atoms close to the scattering resonance.
Within the two channel description introduced above the bound state spectrum of the open channel potential (cf.~Eq.~(\ref{H2B})) is modified by the laser induced coupling to $|\phi_\nu\rangle$.
In the pole approximation, in which only the weakest bound state $|\phi_{-1}^{\rm bg}\rangle$ of $V_{\rm bg}$ is taken into account the interaction picture Hamiltonian (\ref{H2B}) possesses, depending on $I$ and $\Delta$, one or two eigenstates of negative energy.
The normalised dressed states are given by the two-component vector
\begin{equation}
\label{phibound}
  \phi_{\rm d}
  = \left(\begin{array}{c}
        \phi_{\rm d}^{\rm bg} \\
	\phi_{\rm d}^{\rm cl} \end{array}\right)
  = \frac{1}{{\cal N}_{\rm d}}
  \left(\begin{array}{c}
        G_{\rm bg}(E_{\rm d})W\phi_{\nu} \\
	\phi_{\nu} \end{array}\right),
\end{equation}
where $G_{\rm bg}(E_{\rm d})$ is the Greens function of the open channel Hamiltonian evaluated at the in general complex two channel dressed state energy $E_{\rm d}$. 
At time $t=0$, $\phi_\nu$ be normalized to one, such that $\phi_{\rm d}$ is normalized if ${\cal N}_{\rm d}=(1+\langle\phi_{\nu}|W^\dagger G_{\rm bg}(E_{\rm d})^\dagger G_{\rm bg}(E_{\rm d})W|\phi_{\nu}\rangle)^{1/2}$.
One key feature is now that the open channel component significantly contributes not only to $|\phi_{-1}^{\rm bg}\rangle$ away from resonance but also to the near threshold state which causes the resonant behaviour of $a$. 
This is shown in Fig.~\ref{fig:bgfractionofphib}.
The second important property is that this component is, near resonance, far ranged, on the order of the {\it modulus} $|a(\Delta,I)|$ of the scattering length.
To illustrate this we show in Fig.~\ref{fig:phibbgofr} the radial density of the open channel component $\phi_{\rm d}^{\rm bg}(r)$, Eq.~(\ref{phibound}), for peak laser intensities $I_{\rm max}=\,$1.2$\,$kW cm$^{-2}$ and $I_{\rm max}=\,$10$\,$kW cm$^{-2}$, and detunings $\Delta=($2$\pi)\,$192.3$\,$MHz and $\Delta=($2$\pi)\,$1631.2$\,$MHz, respectively, which are chosen such that the real part of $E_{\rm d}$ is just below threshold. (Cf.~also Ref.~\cite{Deb03}.) 

This shows that diatomic correlations may be formed without the necessity that the atoms have to approach each other too closely.
At high intensities the closed channel state contributes less to the dressed state which leads to the saturation of the ground state molecule production through spontaneous decay above ca.~$I=\,$1.0$\,$kW/cm$^2$.  
The difference between $N_{\rm tot}$ and $N_{\rm c}$ as seen in Fig.~\ref{fig:NlossofI} includes atom pairs in the undecayed dressed state as well as in positive energy excited states of their relative motion.
Since the laser intensity is ramped back to zero at the end of the pulse, these atoms are expected to emerge as a burst.

In conclusion we have shown that the atom loss from the condensate fraction due to near resonant photoassociative coupling to an unstable excimer bound state is not limited according to the mean interatomic separation in the sample.
The loss is due to the formation of correlations in elastic two body collisions which is enhanced by means of the long range nature of the near resonant dressed state scaling with the modulus of the scattering length.
Nevertheless the possible yield of ground state molecules is limited and saturates at high intensities.
The time dependence of the loss process may in general not be described by a simple rate equation, in particular within a period after the beginning of the laser pulse which is comparable to the duration of a near resonant elastic collision.

It is proposed to determine the short time loss for significantly larger light intensities as well as the amount of atoms transferred to excited states and bound as ground state molecules. 
    
I thank Thorsten K\"ohler, Paul Julienne, Keith Burnett, Mat\-thi\-as Weidem\"uller, and Olivier Dulieu for many stimulating discussions. This work has been supported through the Deut\-sche Forschungsgemeinschaft.\\[-0.8cm]

\end{document}